\documentclass{amsart}
\usepackage{amsaddr}
\usepackage{geometry}                
\usepackage{graphicx}

\begin{document}

\author{Takumi Kodama}
\address[Shoji Makino]{Life Science Center of TARA, University of Tsukuba, Japan}
\author{Shoji Makino}
\address[Shoji Makino]{Life Science Center of TARA, University of Tsukuba, Japan}
\author{Tomasz M. Rutkowski$^*$}\thanks{$^*$Corresponding author}
\address[Tomasz M. Rutkowski]{Life Science Center of TARA, University of Tsukuba, Japan \\
RIKEN Brain Science Institute, Japan}
\email[Corresponding author]{tomek@bci-lab.info}

\title[Spatial Tactile BCI Paradigm]{Spatial Tactile Brain-Computer Interface Paradigm
 Applying Vibration Stimuli to Large Areas of User's Back}

\maketitle

\markleft{T. KODAMA ET AL.}

\begin{abstract}

We aim at an augmentation of communication abilities of {\it amyotrophic lateral sclerosis} (ALS) patients by creating a {\it brain-computer interface} (BCI) which can control a computer or other device by using only brain activity.
As a method, we use a stimulus--driven BCI based on vibration stimuli delivered via a gaming pad to the user's back. 
We identify P300 responses from brain activity data in response to the vibration stimuli.
The user's intentions are classified according to the P300 responses recorded in the EEG.
From the results of the psychophysical and online BCI experiments, 
we are able to classify the P300 responses very accurately, which proves the effectiveness of the proposed method.
\end{abstract}

\section{Introduction}
\label{sect:introduction}

\begin{figure}[!t]
		\begin{center}
		\includegraphics[width=0.7\linewidth]{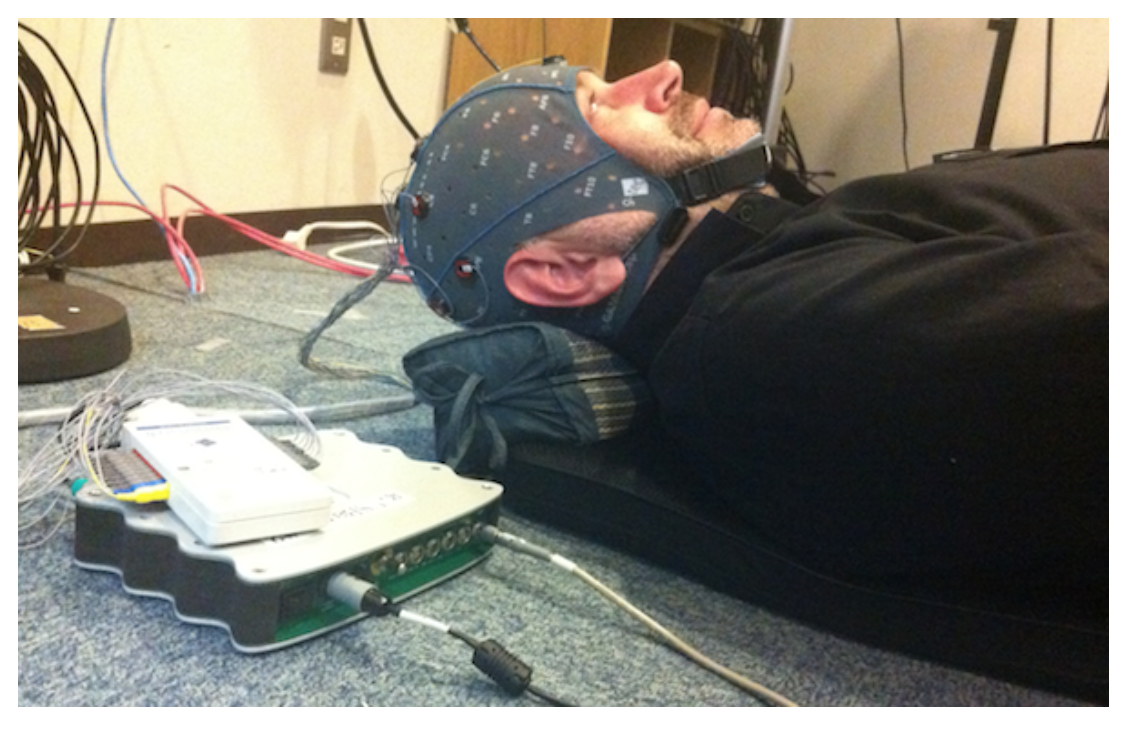}
    	\caption{A btBCI user lying on the gaming pad as in the experiments reported in this paper. The g.USBamp by g.tec with g.LADYbird electrodes is also depicted. The photograph is included with the user's permission.}\label{fig:user}
		\end{center}
\end{figure}

Recently, vibrotactile--based somatosensory modality BCIs have gained in popularity~\cite{MOR,tomekBCI2014}.
We propose an alternative tactile BCI which uses P300 brain responses to a somatosensory stimulation delivered to larger areas of the user's back~\cite{takumiSOTSURON2014}, defined as a back--tactile BCI (btBCI).
We conduct experiments by applying vibration stimuli to the user's back, which allows us to stimulate places at larger distances on the body.
The stimulated back areas are both shoulders, the waist and the hips. In order to do so, we utilize a haptic gaming pad ``ZEUS VYBE'' by Disney \& Comfort Research. An audio signal pad's input allows for the delivery of a sound pattern activating spatial tactile patterns of vibrotactile transducers embedded within the device.
In the experiments reported in this paper, the users lay down on the gaming pad and interacted with tactile stimulus patterns delivered in an oddball--style paradigm to their backs, as shown in Figure~\ref{fig:user}.
The reason for using the horizontal position of the gaming pad, developed for a seated setting, is that bedridden users could easily utilize it and it could also serve as a muscle massage preventing the formation of bedsores. 

The rest of the paper is organized as follows. First we introduce the btBCI experimental protocols and methods. Next the experiment results are discussed. Finally, there is discussion and conclusions are drawn.
\begin{figure}[!b]	
 		\begin{center}
	\includegraphics[width=0.7\linewidth]{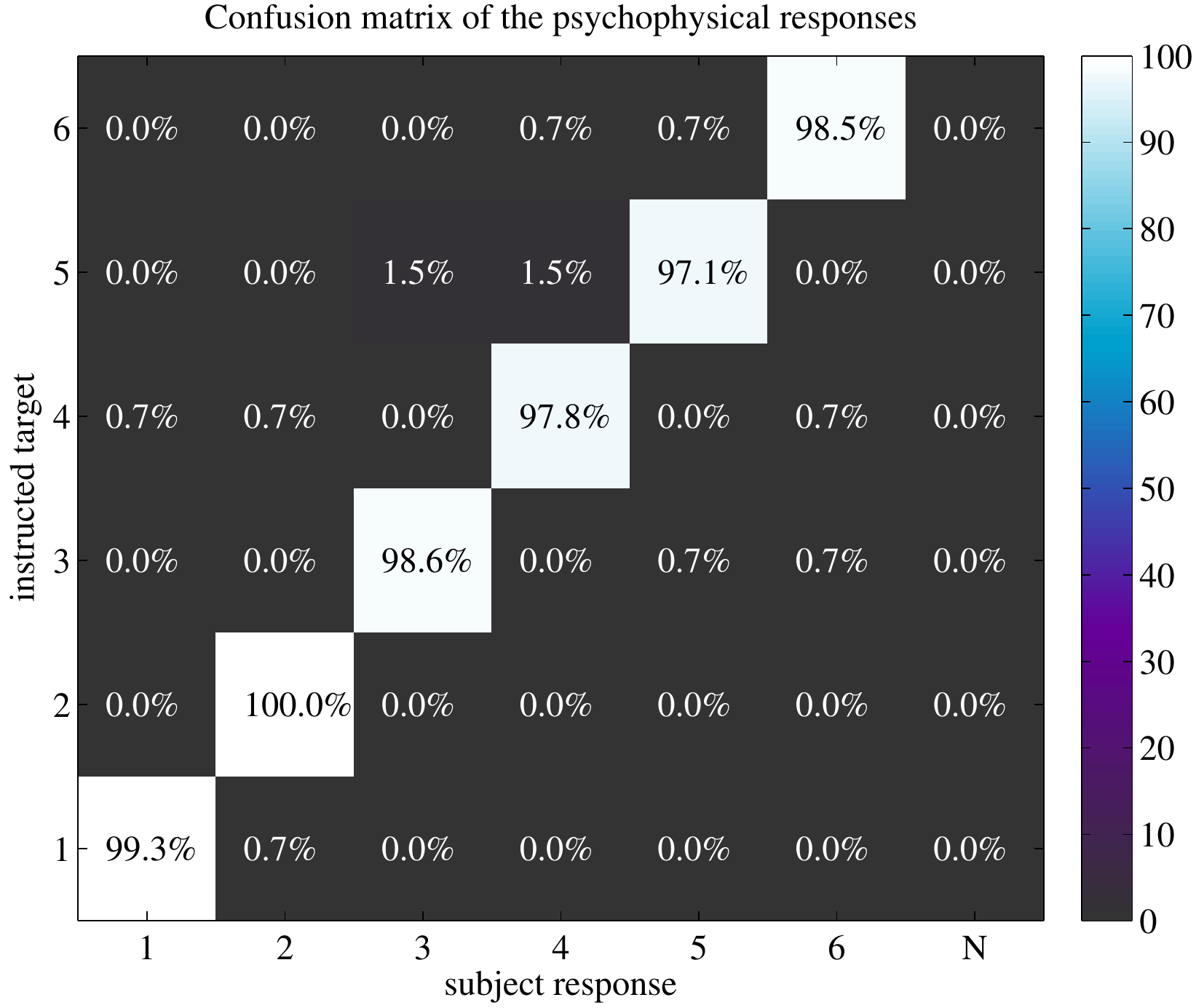}
    \caption{Psychophysical experiment grand mean averaged accuracies with scores above $97\%$ for each of the six commands. ``N'' stands for no--response cases.}\label{fig:confMX}
		\end{center}
\end{figure}

\section{Methods}
\label{sect:methods}
In the research project reported in this paper the psychophysical and online EEG experiments were carried out with able--bodied, BCI--naive users.
Seven healthy users participated in the study (three males and four females) with a mean age of 25 years (standard deviation of 7.8 years). The users were paid for their participation. All the experiments were performed at the Life Science Center of TARA, University of Tsukuba, Japan. 
The online EEG BCI experiments were conducted in accordance with {\it The World Medical Association Declaration of Helsinki - Ethical Principles for Medical Research Involving Human Subjects}. The procedures for the psychophysical experiments and EEG recordings for the BCI paradigm were approved by the Ethical Committee of the Faculty of Engineering, Information and Systems at the University of Tsukuba, Tsukuba, Japan. Each participant signed to give informed consent to taking part in the experiments. 
The psychophysical experiments were conducted to investigate the recognition accuracy and response times to the stimuli delivered from the gaming pad. The behavioural responses were collected as keyboard button presses after instructed targets. In the psychophysical experiment, each single trial was comprised of a randomly presented single target and five non--target vibrotactile stimuli ($120$--targets and $600$--non--targets in a single session). The stimulus duration was set to $300$~ms and the inter--stimulus--interval (ISI) to $700$~ms.
In the btBCI online experiments, the EEG signals were captured with a bio--signal amplifier system g.USBamp by g.tec Medical Instruments, Austria. 
Active EEG electrodes were attached to the sixteen locations {\it Cz, Pz, P3, P4, C3, C4, CP5, CP6, P1, P2, POz, C1, C2, FC1, FC2 and FCz}, as in $10/10$ international system~\cite{TEN}.
A reference electrode was attached to the left mastoid, and a ground electrode to the forehead at the {\it FPz} position. The EEG signals were captured and classified by BCI2000 software~\cite{BCI2000} using a stepwise linear discriminant analysis (SWLDA) classifier~\cite{SWL}.
In each trial, the stimulus duration was set to $250$~ms and the ISI to random values in a range of $350 \sim 370$~ms in order to break rhythmic patterns of presentation. 
The vibrotactile stimuli in the two experimental settings above were generated using the same \emph{MAX~6} program, and the trigger onsets were generated by \emph{BCI2000} EEG acquisition and ERP classification software~\cite{BCI2000}.

\section{Results}
\label{sect:result}

In this section, we report and discuss the results of the psychophysical and btBCI EEG experiments conducted with seven healthy users. 
The psychophysical experiment results are summarized in the form of a confusion matrix depicted in Figure~\ref{fig:confMX}, where the behavioural response accuracies to instructed targets and marginal errors are depicted together with no--response errors, which were not observed with the users participating in our experiments. The grand mean behavioural accuracies were above $97\%$, which proved the easiness of the back vibrotactile stimuli discrimination.
The behavioural response times did not differ significantly as tested with the Wilcoxon rank sum test for medians, which further supported the choice of the experiment set-up with vibrotactile stimuli to the back.
\begin{figure}[!t]
\begin{center}
		\includegraphics[width=\linewidth]{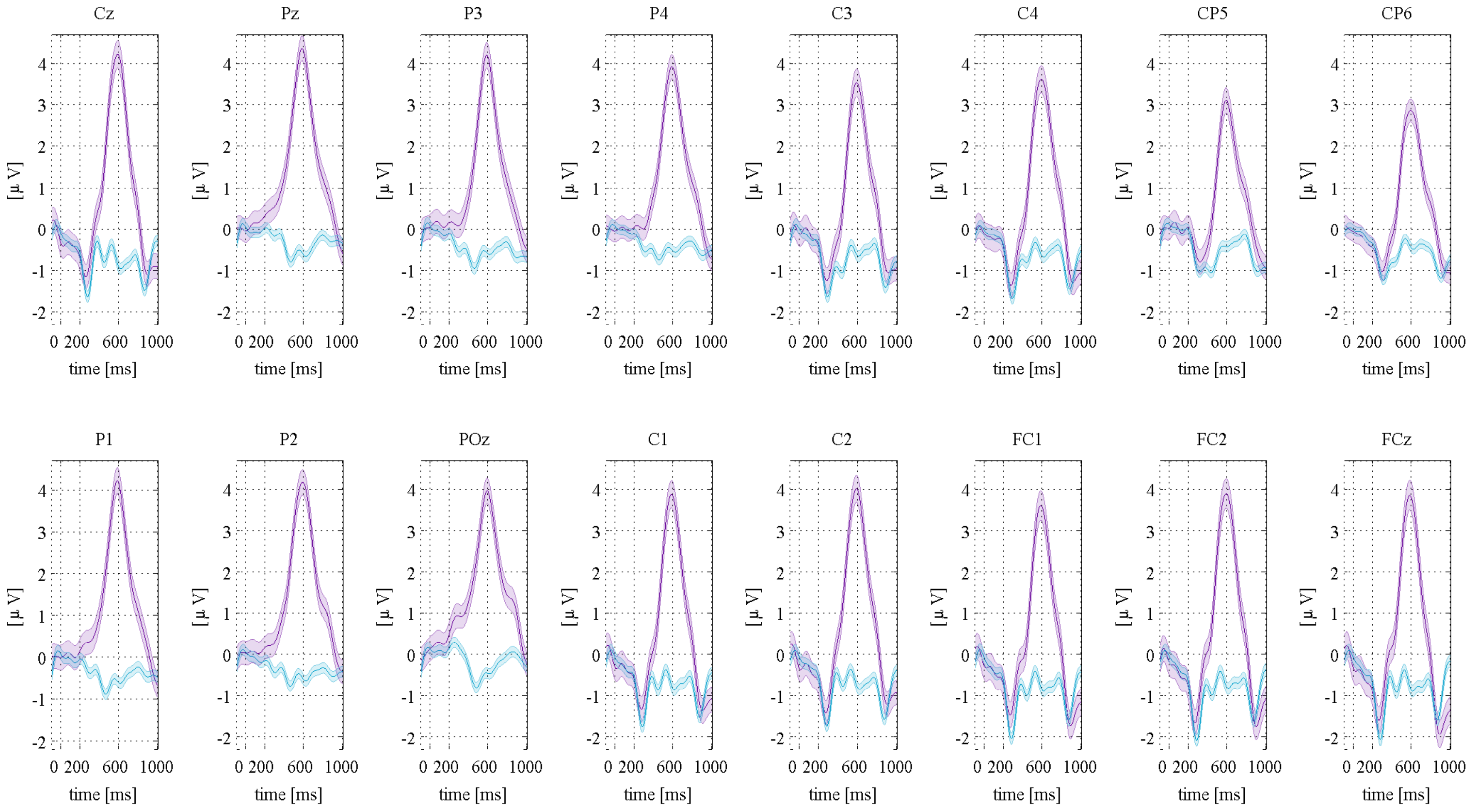}
		\caption{Grand mean averaged ERP for target (purple line) and non--target (blue line) stimuli. The very clear P300 responses are easy to notice for each EEG electrode in the latencies of $200\sim1000$~ms. Eye blinks were rejected in the process of creation of this figure, with a threshold of $80~\mathtt{\mu V}$.\label{fig:allERP}}
\end{center}
\end{figure}

\begin{figure}[!t]
  	\begin{center}
    \includegraphics[width=\linewidth]{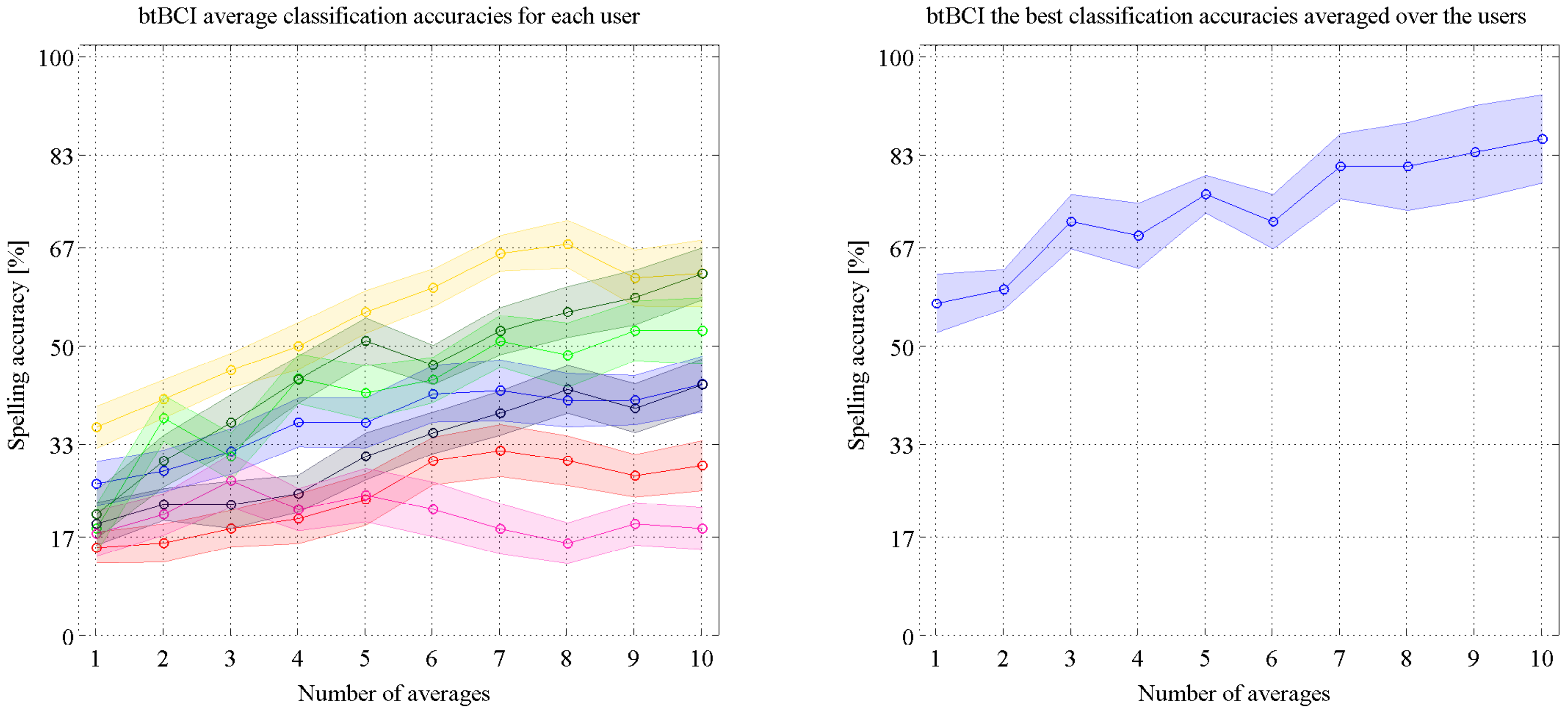}
    \caption{The averaged btBCI accuracies as obtained with a SWLDA classifier with different numbers of averaged ERPs. The left panel presents mean accuracies for each subject separately, together with standard error bars. The right panel presents the averaged maximum scores of all the subjects together.\label{fig:classACC}}
	\end{center}
\end{figure}

The EEG experiment results are summarized in Figures~\ref{fig:allERP}~and~\ref{fig:classACC} in the form of grand mean averaged ERPs and classification accuracies. The grand mean averaged ERPs resulted in very clear P300 responses in latency ranges of $200 \sim 1000$~ms. The SWLDA classification results in online btBCI experiments of six digit spelling are shown in Figure~\ref{fig:classACC}, depicting each user's averaged scores in a range of $16.7\% \sim 62.45\%$ and the best grand mean results in the range of $57.26\% \sim 85.71\%$, both as a function of various ERP averaging scenarios. The chance level was $16.7\%$. The best mean scores show very promising patterns for possible improvements based on longer user training.

\section{Conclusions}
\label{sect:conclusion}
This paper reports results obtained with a novel six--command--based btBCI prototype developed and evaluated in experiments with seven BCI--naive users.   
The experiment results obtained in this study confirm the general validity of the btBCI for six command--based applications and the possibility to further improve the results, as illuminated by the best mean accuracies achieved by the users.

The EEG experiment with the prototype confirms that tactile stimuli to large areas of the back can be used to spell six-digit (command) sequences with mean information--transfer--rates ranging from $0.6$~bit/min to $3.3$~bit/min for $10$--trials averaging based SWLDA classification to $0.5$~bit/min to $10.9$~bit/min for single trial cases. 

The results presented offer a step forward in the development of somatosensory modality neurotechnology applications. Due to the still not very satisfactory interfacing rate achieved in the case of the online btBCI, the current prototype obviously requires improvements and modifications. These requirements will determine the major lines of study for future research. However, even in its current form, the proposed btBCI can be regarded as a possible alternative solution for locked--in syndrome patients, who cannot use vision or auditory based interfaces due to sensory or other disabilities. 

\section*{Author contributions} 
Designed and performed the EEG experiments: TK, TMR. Analyzed the data: TK, TMR. Conceived the concept of the AUTD--based BCI paradigm: TMR. Supported the project: SM. Wrote the paper: TK, TMR.

\section*{Acknowledgements}

The presented research was supported in part by the Strategic Information and Communications R\&D Promotion Program (SCOPE) no.~$121803027$ of The Ministry of Internal Affairs and Communications in Japan.

%

\newcommand{\etalchar}[1]{$^{#1}$}

\end{document}